# GADA: Graph Attention-based Detection Aggregation for Ultrasound Video Classification

Li Chen   Naveen Balaraju   Jochen Kruecker   Balasundar Raju   Alvin Chen

Philips

li.chen_1@philips.com

## Abstract

*Medical ultrasound video analysis is challenging due to variable sequence lengths, subtle spatial cues, and the need for interpretable video-level assessment. We introduce **GADA**, a **Graph Attention-based Detection Aggregation** framework that reformulates video classification as a graph reasoning problem over spatially localized regions of interest. Rather than relying on 3D CNNs or full-frame analysis, GADA detects pathology-relevant regions across frames and represents them as nodes in a spatiotemporal graph, with edges encoding spatial and temporal dependencies. A graph attention network aggregates these node-level predictions through edge-aware attention to generate a compact, discriminative video-level output. Evaluated on a large-scale, multi-center clinical lung ultrasound dataset, GADA outperforms conventional baselines on two pathology video classification tasks while providing interpretable region- and frame-level attention.*

## 1. Introduction

Ultrasound is a widely used, non-invasive, real-time medical imaging modality. Unlike static imaging, ultrasound is inherently dynamic and video-based, with diagnostic cues often emerging through temporal patterns observed across frames. Video-level interpretation is therefore critical for detecting transient or visually subtle pathologies. For instance, pulmonary consolidation may appear as hypoechoic regions with intermittent features such as air bronchograms, while pleural effusion presents as anechoic fluid collections that shift with movement and may be visible in only a few frames. These pathologies are defined by faint, localized features that require temporal continuity for accurate diagnosis. Consequently, robust interpretation requires methods that jointly capture spatial detail and temporal dynamics.

While deep learning has advanced the state-of-the-art in medical image analysis [1], video-level classification for ultrasound remains underexplored. Approaches based on 3D convolutional neural networks (CNNs) [2] face challenges such as high computational cost, fixed-length video requirements, and limited interpretability. A more efficient alternative uses 2D CNNs for frame classification, followed by aggregation of predictions across frames [3]–[5]. However, such methods may fail to capture temporal dependencies and may struggle to distinguish informative frames from noisy or redundant ones.

To better leverage the spatiotemporal structure present in ultrasound videos, two main research directions have emerged. The first emphasizes **temporal attention** by identifying and weighting diagnostically relevant frames. For example, KGA-Net [6] employs attention mechanisms to highlight salient clinical content while filtering out irrelevant frames. The second approach emphasizes **spatial attention** by incorporating region-level detection within frames. These methods [7] localize pathology-relevant areas and aggregate region-level cues across time, enabling more focused and interpretable predictions.

While both strategies improve performance and interpretability, they are typically applied in isolation. Graph Neural Networks (GNNs) offer a natural framework for structured reasoning in videos by representing detected regions as nodes and linking them across frames based on spatial proximity and temporal adjacency. Their ability to integrate features across non-uniform spatiotemporal contexts makes them well suited for ultrasound. Among GNN architectures, graph attention networks are particularly attractive for detection-driven medical video classification tasks because they assign relative importance among spatial–temporal relationships while maintaining interpretability through attention weights.

This work introduces Graph Attention-based Detection Aggregation (GADA), a framework that integrates spatial and temporal attention for ultrasound video classification (Figure 1). GADA constructs a spatiotemporal graph, where nodes represent region-level detections and edges capture spatial proximity and temporal continuity. A graph attention network aggregates these features through edge-aware message passing, producing an interpretable and adaptive video-level prediction. GADA handles variable-length inputs and detection densities while enabling integrated spatiotemporal reasoning over detected regions.

Table 1. Comparison of ultrasound video classification methods. Each method is evaluated based on key criteria: spatial attention (focus on pathology-relevant regions), temporal modeling (ability to capture temporal dependencies), temporal attention (identification of key frames), interpretability (level of diagnostic insight provided), and major limitations.

| Method | Use spatial attention | Learn temporal knowledge | Use temporal attention | Interpretability | Main limitations |
|---|---|---|---|---|---|
| 3D spatiotemporal convolutions [8] | No | Yes | No | None | High computational cost; fixed-length inputs |
| 2D CNN + frame averaging [3] | No | No | No | Frame-level | Ignores temporal structure; sensitive to noise |
| 2D CNN + LSTM [4] | No | Yes | No | None | Slow training; sensitive to noise; limited interpretability |
| KGA-Net [6], Key-frame guided network [10] | No | Yes | Yes | Frame-level | Operates on full-frame features; may miss localized pathology |
| Detection [19] + frame averaging | Yes | No | No | Region-level (per frame) | No temporal modeling; lacks frame linkage |
| Tracklets + LSTM [7] | Yes | Yes | No | Region-level (along tracklet) | Limited temporal reasoning; tracking is error-prone |
| Frame graph GNN [13] | No | Yes | Yes | Frame-level | Does not explicitly model spatial detection regions |
| GADA (proposed) | Yes | Yes | Yes | Region-level (per frame) | - |

The key contributions of this work are as follows:

- We propose a novel detection-driven approach for ultrasound video classification that integrates region-level predictions to jointly model spatial and temporal information.

- We develop a graph attention network to dynamically aggregate variable numbers of detections across frames, enabling flexible and interpretable video-level predictions.

- We evaluate our method on a large, multi-center ultrasound dataset, showing superior performance compared to conventional baselines, including frame averaging, LSTMs, and tracking-based methods.

## 2. Related Work

### 2.1. Ultrasound Video Classification

Ultrasound imaging is inherently dynamic, requiring interpretation across multiple frames to capture temporal patterns. Traditional approaches often adapt general video classification models such as 3D CNNs [2]. For instance, EchoNet-Dynamic [8] uses spatiotemporal convolutions to predict ejection fraction from echocardiography videos. While effective, 3D CNNs are computationally intensive and constrained to fixed-length inputs, limiting their practicality for variable-length ultrasound videos.

An alternative approach is to extract frame-level features or generate frame-level predictions using 2D CNNs and apply temporal aggregation. This can be done using simple rule-based methods (e.g., averaging) or learnable models such as long short-term memory (LSTM) networks [9]. For example, Naser et al. [3] applied a 2D CNN to classify individual frames of echocardiographic videos and then averaged the predictions to yield a video-level classification. Similar strategies have been used in breast nodule analysis [4] and fetal plane classification [5], combining 2D CNNs with LSTMs for temporal modeling. However, these methods have key limitations: frame averaging ignores temporal dependencies, and sequential models like LSTMs are slow and sensitive to noisy or redundant frames. Moreover, they often lack mechanisms to prioritize diagnostically relevant content.

### 2.2. Video Analysis with Temporal Attention

To address these issues, attention-based models have been proposed to emphasize diagnostically meaningful frames. KGA-Net [6], for example, learns attention weights over key frames to emphasize salient lesion appearances in fetal ultrasound. Similarly, [10] introduces motion-aware attention to guide thyroid nodule classification by highlighting diagnostically significant frames, mimicking a clinician's focus during scanning. Although these approaches improve frame selection, they still rely on full-frame features and may miss subtle or spatially localized pathologies.

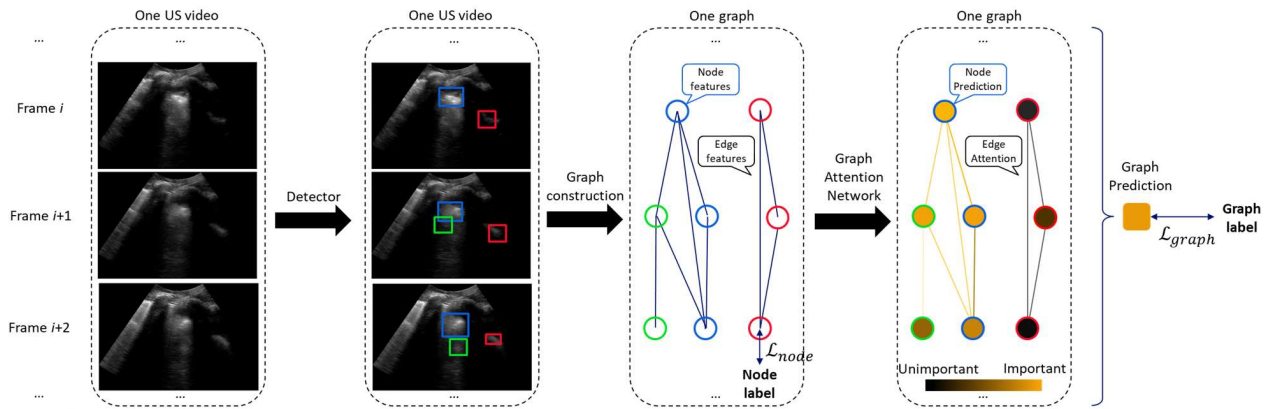

Figure 1. Overview of the proposed method for lung consolidation classification. Frame-level detector outputs are used to construct a graph for each video, where nodes represent detected regions and edges capture spatiotemporal relationships. A graph attention network predicts node-level scores and assigns attention weights to edges, which are then aggregated into a video-level prediction.

### 2.3. Video Analysis with Spatial Attention

Beyond frame-level attention, recent studies have introduced spatial attention and detection-based pipelines to incorporate localization of pathology-relevant regions. For example, Kerdegari et al. [11] propose a spatial attention module that generates importance masks to detect B-line artifacts in lung ultrasound. TASL-Net [12] uses spatial attention to identify early perfusion changes in contrast-enhanced ultrasound. Specifically, they use a structure similarity-based early enhancement detector to guide attention to relevant perfusion changes. This design improves both interpretability and accuracy.

Li et al. [7] propose a detection-based framework that tracks regions of interest (ROIs) across frames to form tracklets, which are then processed by a CNN-LSTM classifier. The final prediction is based on the most confident tracklet. While this method improves spatial focus, it introduces additional complexity through temporal tracking and lacks robust modeling of long-range temporal dependencies.

### 2.4. Video Analysis with Graph Neural Networks

Graph neural networks (GNNs) have recently gained traction for modeling structured spatial-temporal data. EchoGNN [13], for instance, builds a graph where each node corresponds to a frame in an echocardiogram video, with edges encoding temporal dependencies for predicting ejection fraction. However, it does not explicitly model spatial relationships between detected regions.

GNNs have also been applied to static ultrasound images. In [14], nodes represent individual images and edges capture feature similarity for breast tumor detection. However, these approaches are designed for static settings and do not leverage the spatiotemporal information critical for video-based classification.

### 2.5. Graph Attention for Spatiotemporal Aggregation

General GNN literature has explored combining graph attention with message passing to better handle node and edge information in dynamic settings [15]. Our work builds on this by constructing a sparse, detection-driven spatiotemporal graph tailored for ultrasound video classification. Nodes represent detected regions, and edges encode spatial proximity, temporal adjacency, and feature similarity. A graph attention network dynamically aggregates node-level predictions. Unlike prior work, our method directly integrates detection and video classification through edge-aware attention, achieving both adaptability to variable-length videos and improved interpretability.

A summary comparison of related methods is provided in Table 1.

## 3. Methods

We propose **GADA** (Graph Attention-based Detection Aggregation), a detection-driven, graph-based framework for ultrasound video classification. The approach consists of two stages: detecting pathology-relevant regions at the frame level, followed by spatiotemporal aggregation using a graph attention network.

## 3.1. Frame-Level Pathology Detection

We first train a frame-level object detector $D$ to localize pathological regions. For each frame $f_t$ in a video $V = \{f_1, f_2, ..., f_T\}$, the detector outputs a variable set of bounding boxes:
$$\mathcal{D}(f_t) = \{b_t^{(i)} = (x, y, w, h, c)^{(i)}\}_{i=1}^{N_t}$$
where $(x, y, w, h)$ denote normalized bounding box parameters, $c$ is the detection confidence, and $N_t$ is the number of boxes detected in frame $t$.

To construct graphs, we empirically set a low minimum confidence threshold of $c > 0.01$ to ensure nearly all the positive videos contribute at least one detection.

## 3.2. Graph Construction

Each ultrasound video is represented as a single spatiotemporal graph $G = (V, E)$, where:

**Nodes** $V$ correspond to individual detection boxes $\boldsymbol{b}_t^{(i)}$. Each node is associated with a feature vector:
$$\boldsymbol{b}_t^{(i)} = (x, y, w, h, c)^{(i)}$$

Node Labels $\hat{y}_v \in \{-1, 1\}$ are derived from the maximum Intersection over Union (IoU) with ground truth boxes on the frame $\mathcal{G}$:
$$\hat{y}_v = \begin{cases} 1, & \text{if } \max_{g \in \mathcal{G}} \text{IoU}(b_v, g) > 0 \\ -1, & \text{otherwise} \end{cases}$$

**Edges** $E$ connect boxes in adjacent frames (within a $\epsilon = \pm 5$-frame window) if their IoU exceeds a threshold $\delta$. Between the pair of nodes $u$ and $v$, each edge $e = (u, v)$ is assigned with a feature vector:
$$\boldsymbol{e}_{uv} = [\text{IoU}(\boldsymbol{b}_u, \boldsymbol{b}_v), |\boldsymbol{c}_u - \boldsymbol{c}_v|_2, |t_u - t_v|]$$
where $\boldsymbol{c}_u$ is the box center and $t_u$ is the frame index of node $u$.

The graph is sparse, and both node and edge attributes are used during graph reasoning.

## 3.3. Graph Attention Network for Feature Aggregation

To aggregate region-level information across frames, we adopt a Message Passing Graph Transformer, inspired by [15]. Each graph is passed through a stack of $L = 3$ Graph Transformer layers, each with $H = 4$ attention heads and hidden dimension of 64.

**Message Passing:** During message passing, each node $v$ aggregates information from its neighbors to update its own representation, illustrated in Figure 2. For node $v$, the updated feature at layer $l + 1$ is:

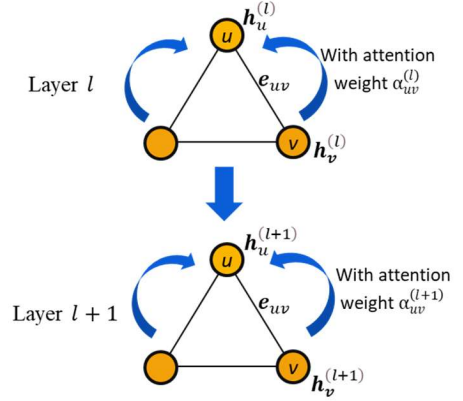

Figure 2. The message passing framework used in GADA.

$$\boldsymbol{h}_u^{(l+1)} = W \cdot \boldsymbol{h}_u^{(l)} + \sum_{u \in \mathcal{N}(v)} \alpha_{uv}^{(l)} \cdot \phi(\boldsymbol{h}_v^{(l)}, \boldsymbol{e}_{uv}),$$
$$\boldsymbol{h}_u^{(0)} = \boldsymbol{b}_t^{(i)}$$
where $W$ is a learnable linear transformation, $\phi$ is a learnable edge-aware transformation (e.g., MLP), and $\alpha_{uv}^{(l)}$ is the attention coefficient computed via multi-head scaled dot-product attention:

$$\alpha_{uv}^{(l)} = \text{softmax}_u \left( \frac{(\boldsymbol{Q}_v^{(l)})(\boldsymbol{K}_u^{(l)})^\top}{\sqrt{d}} + \psi(\boldsymbol{e}_{uv}) \right)$$

Here, $\boldsymbol{Q}, \boldsymbol{K}$ are learned projections, $d$ is the attention dimension, and $\psi$ is a scalar edge embedding network.

**Graph Level Prediction:** Given the attention coefficient from node $u$ to node $v$, the node weight $\beta_v$ is computed as the weighted incoming attention:
$$\beta_v = \sum_{u \in \mathcal{N}(v)} \alpha_{uv}$$
where $\mathcal{N}$ indicates the connected nodes from node $v$.

The final graph-level prediction $y_G \in \{-1, 1\}$ is the weighted sum of all node-level predictions:
$$y_G = \sum_{v \in V} \beta_v \cdot y_v$$

## 3.4. Training Objective

We jointly optimize the node and graph classification losses. The total loss is:
$$\mathcal{L} = \mathcal{L}_{node} + \mathcal{L}_{graph}$$
**Node Loss** $\mathcal{L}_{node}$: A mean squared error (MSE) between predicted node outputs and binarized node labels

$\hat{y}_v$ (1 for any IoU matches, -1 otherwise), encouraging per-region discrimination.

$$\mathcal{L}_{node} = \frac{1}{|V|} \sum_{v \in V} (\hat{y}_v - y_v)^2$$

**Graph Loss** $\mathcal{L}_{graph}$: Binary cross-entropy with logits between the graph-level prediction and the ground truth video label $\hat{y}_G$ (0 for negative, 1 for positive).

$$\mathcal{L}_{grap} = -y \log(\sigma(\hat{y}_G)) - (1-y) \log(1 - \sigma(\hat{y}_G))$$

where $\sigma$ indicates a sigmoid function.

## 4. Results

### 4.1. Dataset and Experimental Methods

We evaluated our method on a multi-center clinical dataset of lung ultrasound videos collected from nine medical centers across the United States. The dataset comprises a diverse range of clinical conditions and pathology, as well as variation in ultrasound hardware systems and imaging conditions. Videos were acquired at frame rate of 20 to 30 per second and at least 3 seconds in duration, leading to 60 to 180 frames. Detailed dataset characteristics are summarized in Table 2.

Each video was independently annotated by at least two expert physicians for the presence or absence of pathological findings, specifically lung consolidation and/or pleural effusion, following established clinical guidelines [16], [17]. In cases of disagreement, final binary video-level labels were determined via majority vote among at least three expert physicians. Additionally, frame-level bounding boxes for each pathological finding were annotated by ultrasound-trained researchers and reviewed by at least one physician. All experiments were conducted with subject-level division into training, validation, and test sets. The validation and test sets were balanced with respect to video-level class labels.

### 4.2. Implementation Details

The graph attention network comprises 21,304 learnable parameters and is optimized using Adam [18] with a learning rate of 1×10⁻⁴ over 10,000 epochs. To improve robustness, we trained on graphs generated from detector outputs at varying stages of training, excluding the first 100 epochs when the detector is still significantly undertrained. For each training batch, 100 videos are randomly sampled with balanced positive and negative labels. Graphs are reconstructed every 50 epochs using detection checkpoints sampled every 10 epochs. This strategy introduces variation in detection quality, encouraging the graph model to generalize in the presence of noisy detection results.

We use the widely known YOLO model [19] as the detector in our experiments. While recent state-of-the-art object detection architectures could potentially achieve higher accuracy, we deliberately chose YOLO (a standard baseline detection framework) to emphasize that GADA is detector-agnostic. Additionally, when considering clinical deployment, YOLO is lightweight and well suited to point-of-care ultrasound settings, where computational resources are often limited.

To ensure graph construction for all positive cases, we set a low detection confidence threshold of 0.01. Videos without any detections above this threshold are automatically classified as negative by the graph model.

Binary video classification performance was assessed using the Area Under the Receiver Operating Characteristic Curve (AUC), video level sensitivity and specificity with optimal threshold derived from maximum averaged sensitivity and specificity on the validation set.

### 4.3. Key parameter choices in graph construction

To evaluate the impact of graph construction design on the final classification performance, we conducted an ablation study by varying two key hyperparameters used in graph edge formation: the frame search window size $\epsilon$ and the box IoU threshold $\delta$. The frame search window controls the temporal range of frame-to-frame connections, while the IoU threshold determines whether two detected boxes across frames are similar enough to form an edge. We tested four values for each parameter: $\epsilon \in \{3, 5, 10, 60\}$ and $\delta \in \{0, 0.1, 0.3, 0.5\}$. A larger $\epsilon$ allows long-range temporal dependencies but may introduce noisy associations, while a stricter IoU threshold $\delta$ enforces spatial precision at the cost of graph sparsity. From Table 3 and Table 4, the systematic analysis reveals our choices of $\epsilon = 5$ and $\delta = 0$ achieves the best overall performance, slightly outperforming other configurations.

Table 2: Clinical imaging datasets used in this study.

| Counts (Sites = 9) | Consolidation | | | Pleural Effusion | | |
|---|---|---|---|---|---|---|
| | Train | Val | Test | Train | Val | Test |
| Subjects | 263 | 76 | 78 | 228 | 72 | 76 |
| Videos | 1,578 | 396 | 484 | 1,220 | 236 | 208 |
| Positive videos | 619 | 198 | 242 | 383 | 118 | 104 |
| Frames | 99.8k | 25.0k | 31.4k | 77.8k | 14.3k | 12.6k |

Table 3: Comparison of Frame window $\epsilon$ for video-level classification of lung consolidation and pleural effusion.

| Frame window $\epsilon$ | Consolidation AUC | Pleural effusion AUC |
|---|---|---|
| 3 | 0.942 | 0.952 |
| **5 (proposed)** | **0.943** | **0.960** |
| 10 | 0.939 | 0.959 |
| 60 | 0.916 | 0.931 |

Table 4: Comparison of IoU threshold $\delta$ for video-level classification of lung consolidation and pleural effusion.

| IoU threshold $\delta$ | Consolidation AUC | Pleural effusion AUC |
|---|---|---|
| **0.0 (proposed)** | **0.943** | **0.960** |
| 0.1 | 0.935 | 0.957 |
| 0.3 | 0.938 | 0.945 |
| 0.5 | 0.929 | 0.951 |

Table 5: Comparison of GADA and baseline methods for video-level classification of lung consolidation and pleural effusion. McNemar p-value indicates the significance (if p-value<0.05) when the proposed GADA model compares with baseline methods.

| Methods | Temporal attention | Spatial attention | Consolidation | | | | | Pleural effusion | | | | |
|---|---|---|---|---|---|---|---|---|---|---|---|---|
| | | | AUC | Sens-itivity | Spec-ificity | Acc-uracy | McNemar p-value | AUC | Sens-itivity | Spec-ificity | Acc-uracy | McNemar p-value |
| CNN + Frame Avg | No | No | 0.927 | 0.864 | 0.822 | 0.843 | 0.0422 | 0.945 | 0.625 | **0.990** | 0.808 | 0.0058 |
| Detector + Frame Avg | No | No | 0.919 | 0.831 | 0.806 | 0.818 | <0.0001 | 0.934 | 0.731 | 0.913 | 0.822 | <0.0001 |
| CNN + LSTM [4] | No | No | 0.938 | 0.872 | 0.843 | 0.857 | 0.2353 | 0.943 | 0.625 | **0.990** | 0.808 | 0.0072 |
| KGA-Net [6] | Yes | No | 0.938 | 0.851 | 0.851 | 0.851 | 0.1307 | 0.943 | 0.798 | 0.933 | 0.865 | 0.3613 |
| Tracklets + LSTM [7] | No | Yes | 0.906 | 0.864 | 0.781 | 0.822 | 0.0001 | 0.909 | 0.769 | 0.904 | 0.837 | 0.0310 |
| **GADA (proposed)** | **Yes** | **Yes** | **0.943** | **0.888** | **0.872** | **0.880** | - | **0.960** | **0.875** | 0.913 | **0.894** | - |

### 4.4. Comparison with Existing Methods

We benchmarked GADA against a range of baseline and state-of-the-art methods for video classification:

- **CNN + Frame Averaging:** Frame-level predictions from a 2D CNN are averaged across all frames to produce a video-level score. This simple aggregation allows evaluation of the CNN's standalone performance on video classification.

- **Detector + Frame Averaging:** For each frame, the maximum detection confidence is used as the frame-level prediction, which is then averaged to obtain a video-level score. This approach evaluates the detector's performance on video classification tasks independently and provides a baseline for GADA, which builds on detection outputs.

- **CNN + LSTM** [4]: Frame-level CNN features are passed through an LSTM for temporal aggregation and video-level classification.

- **KGA-Net** [6]: A video classification model using temporal attention to highlight frames with salient clinical findings.

- **Tracklets + LSTM** [7]: Regions of interest are tracked across frames to form tracklets. Box coordinates and confidences along each tracklet are passed through an LSTM, with the maximum tracklet prediction used as the video level score. For fair comparison with GADA, image patches used in [7] were not applied in our experiments.

GADA outperformed all baseline methods (Table 5), with notable improvements over both CNN frame-based methods and detector-based approaches. We assessed the statistical significance of these improvements using McNemar's test [20], [21]. For consolidation video classification, GADA significantly (p<0.05) outperformed three out of five baseline methods (CNN + Frame Avg, Detector + Frame Avg, and Tracklets + LSTM). For pleural effusion video classification, GADA significantly (p<0.05) outperformed all baselines with the exception of KGA-Net.

Although comparisons with other state-of-the-art object detectors are possible, **GADA is detector-agnostic**: it functions as a graph-based aggregation framework that can be applied to the outputs of any detector. Improvements in the underlying detection model can directly enhance overall performance, but the focus of this work is on robust and efficient video-level classification from frame-wise detections, particularly in long, variable-length videos with localized pathology.

### 4.5. Ablation Studies

To assess the contribution of individual input features, we conducted ablation studies by selectively removing components from the node and edge representations (Table 6). Among the features tested, removing detection confidences led to the largest decrease in video classification performance. In contrast, box positions did not contribute to classification performance and were ultimately excluded from the final feature set. Other features, such as box size and edge attributes (e.g., IoU, spatial distance, and frame difference), had moderate impact on classification performance.

Table 6: Ablation study results for video-level classification of lung consolidation and pleural effusion using GADA.

| Box position (x, y) | Box size (w, h) | Box confidence (c) | Edge features | Consolidation, AUC | Pleural effusion, AUC |
|---|---|---|---|---|---|
| ✓ | ✓ | ✓ | ✓ | 0.937 | 0.955 |
|   | ✓ | ✓ | ✓ | **0.943** | **0.960** |
| ✓ | ✓ | ✓ |   | 0.931 | 0.950 |
| ✓ | ✓ |   | ✓ | 0.924 | 0.936 |
| ✓ |   | ✓ | ✓ | 0.938 | 0.946 |
|   | ✓ | ✓ |   | 0.935 | 0.950 |
|   |   | ✓ | ✓ | 0.938 | 0.941 |
|   |   | ✓ | ✓ | 0.935 | 0.950 |

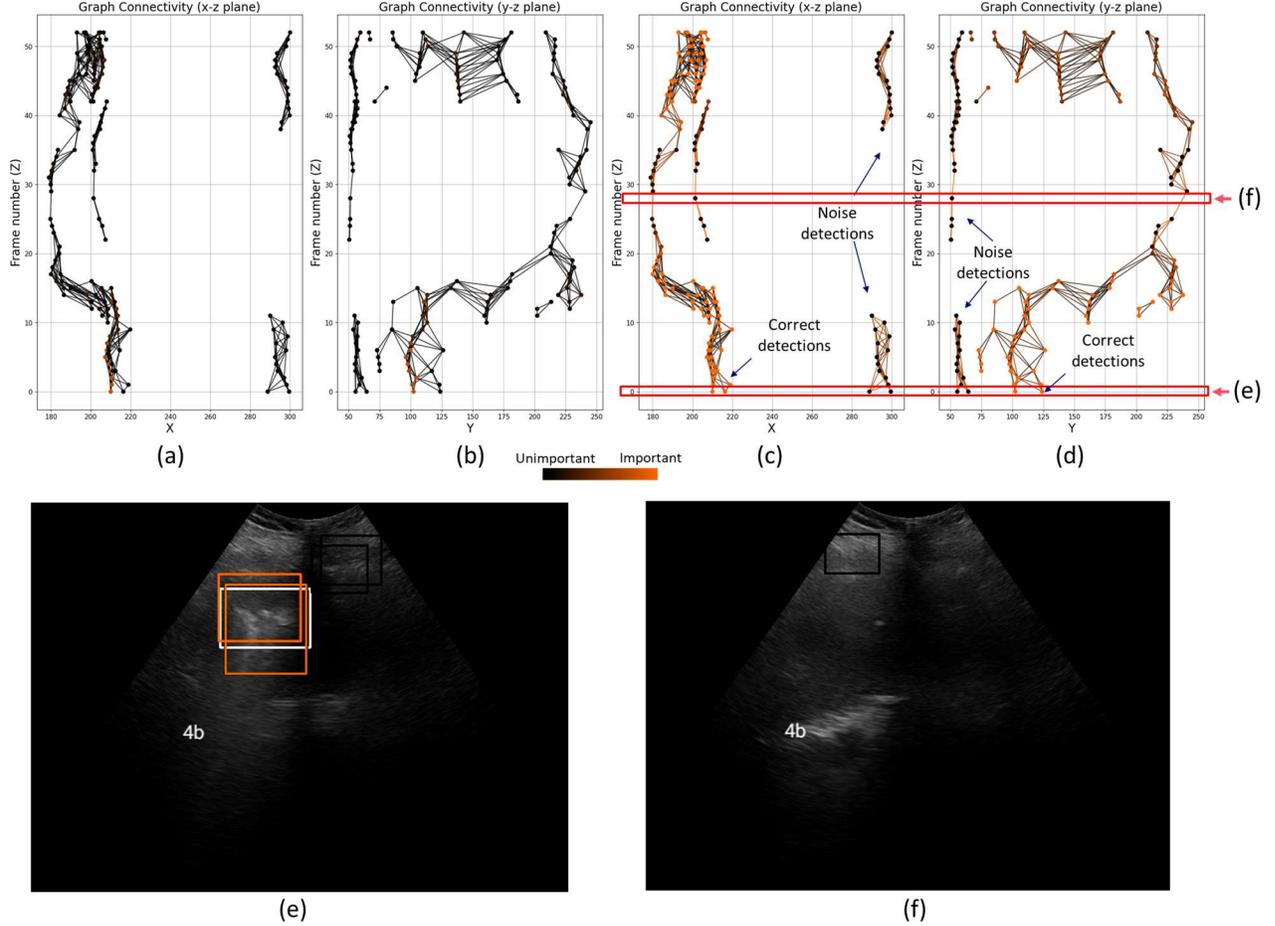

Figure 3. Visualization of node predictions and edge attention before and after GADA processing. (a) and (b) show the graph in the X–Z and Y–Z planes, respectively, before applying the graph attention network. Node color represents initial detection confidence, from black (low) to orange (high). (c) and (d) display the same projections after GADA inference. Node color reflects the predicted probability of overlap with ground truth (i.e., the node-level output) and edge color indicates the learned attention weights. GADA effectively identifies a coherent sequence of high-confidence detections corresponding to pathology while suppressing isolated or noisy nodes. (e) and (f) show example ultrasound frames: the white box indicates ground truth lung consolidation, and colored boxes represent node predictions, with color indicating graph-based importance scores. GADA correctly highlights the true pathology in (e) and down-weights a spurious detection in (f).

### 4.6. Impact of Detector Quality on GADA Performance

One potential concern is that GADA's performance depends on the quality of the object detector, since detected boxes are the source for graph construction. To evaluate robustness to detector quality, we conducted an additional experiment in which the trained GADA weights were fixed and input detections were replaced with outputs from different checkpoints of the same YOLO detector, ranging from optimal to substantially suboptimal. Specifically, we selected checkpoints corresponding to the optimal detector model, as well as models with $\pm 1$, $\pm 10$, and $\pm 50$ epochs from the optimal training point.

Video-level AUC remained highly stable across all detector checkpoints. The maximum difference in AUC compared to the best-detector case was 0.0025, indicating that minor or moderate variations in detection quality had minimal impact on final video-level classification. These results suggest that GADA's graph aggregation mechanism is robust to imperfect detections and can mitigate error propagation from the detection stage.

### 4.7. Visualization of Graph Predictions

The predicted node outputs and edge attention scores from GADA offer insight into how the model integrates spatial and temporal information across a video. As illustrated in Figure 3, high-confidence detections initially appear sporadically, with some clusters likely representing noise. After applying GADA, the model identifies a coherent sequence of detections corresponding to true pathology, characterized by elevated node scores and strong edge attention. Irrelevant detections, in contrast, receive low node scores and weak edge weights, reducing their impact on the final prediction. GADA's explicit node- and edge-level outputs enable visualization of the detections and spatiotemporal relationships driving the final video-level prediction – potentially supporting interpretability and trust in clinical decision-making.

## 5. Conclusions

This work introduced GADA, a Graph Attention-based Detection Aggregation framework for video classification. GADA represents each video as a spatiotemporal graph, with nodes corresponding to detected pathological regions and edges capturing spatial and temporal relationships. A graph attention network aggregates region-level predictions, enabling the model to handle variable-length inputs and focus on clinically meaningful cues. Compared to pooling, sequential models (e.g., LSTMs), and handcrafted aggregation heuristics, GADA achieves superior performance while maintaining interpretability through attention-weighted contributions from specific regions and frames.

While the current implementation trains the detector and graph model separately and uses fixed criteria for graph construction, future improvements could include end-to-end optimization and learnable graph connectivity. Additionally, the current node and edge features were selected empirically; incorporating richer or domain-informed features, as well as integrating global graph-level attributes such as imaging parameters or patient-level information, may further improve performance. Overall, GADA offers a robust, interpretable, and adaptable framework for ultrasound video classification and holds promise for broader applications across imaging modalities and clinical diagnostic tasks.

## Acknowledgement

This study was funded in part by the U.S. Department of Health and Human Services (HHS); Administration for Strategic Preparedness and Response (ASPR); Biomedical Advanced Research and Development Authority (BARDA), under contract number 75A50120C00097. The contract and federal funding are not an endorsement of the study results, product or company. The findings and conclusions in this report are those of the author(s) and do not necessarily represent the views of the Department of Health and Human Services or its components.

We gratefully acknowledge the following individuals for their contributions to clinical data acquisition, curation, and annotation: Christopher L. Moore, Cristiana Baloescu, Ryan Denkewicz, Mina Hesami, Bryson Hicks, Nikolai Schnittke, Maria Parker, Aaron Silver, Michael Petrovich, Alfredo Sabbaj, Lianjun Hu, Daniela Chan, Meihua Zhu, Matthew Kaili, Caelan Thomas, Maggie Feuerherdt, Yuan Zhang, Cynthia R. Gregory, Kenton W. Gregory, David O. Kessler, Di Coneybeare, Laurie Malia, Jeffrey Shupp, Courosh Mehanian, Sourabh Kulhare, and Rachel Millin.